%% file: main.tex
\documentclass[11pt]{article} % For LaTeX2e
\usepackage{smile}
\usepackage{epstopdf}
\usepackage{wrapfig}
\usepackage[colorlinks, linkcolor=blue, anchorcolor=blue, citecolor=blue]{hyperref}
\usepackage[margin=1in]{geometry}
\usepackage[normalem]{ulem}
\usepackage[export]{adjustbox} % http://ctan.org/pkg/adjustbox
\usepackage{mathtools, cuted}
\usepackage{natbib}
\usepackage{enumerate}
\usepackage{enumitem}

\usepackage{pifont}
\usepackage{multirow}
\usepackage{caption}
\usepackage{subcaption}
\usepackage{enumitem}
\usepackage{tabularx}  % This package is used for controlling the table width
\usepackage{multirow}  % This package is used for spanning rows
\usepackage{booktabs}  % This package is used for formatting tables
\newcommand{\cmark}{\ding{51}}
\newcommand{\xmark}{\ding{55}}
\newcommand{\ours}{\text{DeepTagger}}

\usepackage{kpfonts}
\DeclareMathAlphabet{\mathsf}{OT1}{cmss}{m}{n}

\SetMathAlphabet{\mathsf}{bold}{OT1}{cmss}{bx}{n}

\providecommand{\norm}[1]{\|#1\|}

\newcommand{\commentout}[1]{}
\newtheorem*{theorem*}{Theorem}

\title{\bf DeepTagger: Knowledge Enhanced Named Entity Recognition for Web-Based Ads Queries}

\author{
Simiao Zuo\thanks{Corresponding author.}, Pengfei Tang, Xinyu Hu, Qiang Lou, Jian Jiao, Denis Charles \\
\texttt{\{simiaozuo,pengfeitang,xinyuhu,qilou,jian.jiao,cdx\}@microsoft.com} \\
Microsoft
}
\date{}

\begin{document}

\maketitle

\begin{abstract}
\noindent
Named entity recognition (NER) is a crucial task for online advertisement. State-of-the-art solutions leverage pre-trained language models for this task. However, three major challenges remain unresolved: web queries differ from natural language, on which pre-trained models are trained; web queries are short and lack contextual information; and labeled data for NER is scarce. We propose DeepTagger, a knowledge-enhanced NER model for web-based ads queries. The proposed knowledge enhancement framework leverages both model-free and model-based approaches.
For model-free enhancement, we collect unlabeled web queries to augment domain knowledge; and we collect web search results to enrich the information of ads queries. We further leverage effective prompting methods to automatically generate labels using large language models such as ChatGPT.
Additionally, we adopt a model-based knowledge enhancement method based on adversarial data augmentation. We employ a three-stage training framework to train DeepTagger models. Empirical results in various NER tasks demonstrate the effectiveness of the proposed framework.
\end{abstract}

\input{0-introduction}
\input{0-data-augmentation}

\input{0-adv-augmentation}

\input{0-training}
\input{0-data-preparation}

\input{0-experiments}
\input{0-related-works}
\input{0-conclusion}

\bibliography{main}
\bibliographystyle{ims}

\end{document}

%% file: 0-introduction.tex
\section{Introduction}

Named Entity Recognition (NER) is the task of classifying each token in an input sequence into predefined categories. For example, in a query such as ``hotels in Seattle'', we need to identify ``\textit{hotels}'' as a type of product and ``\textit{Seattle}'' as a location designator. In the advertisement domain, NER lays the foundation for subsequent applications such as  product retrieval \citep{cheng2021end}, query rewriting \citep{wen2019building}, and attribute value extraction \citep{zhang2021queaco}. Existing works leverage pre-trained language models (PLMs \citep{devlin2018bert}) for NER. These models are pre-trained on large natural language corpora and contain rich syntactic and semantic information.

There are three major challenges when applying PLMs to NER for web-based ads queries. First, there is a domain shift between web queries and natural language. Most web queries lack grammatical components such as verbs and subjects. For example, ``home furniture bedroom'' is an informative web query but not a grammatically correct sentence. Additionally, web queries involve specification attributes and product models that are uncommon in natural language. These properties create a domain shift that hinders the performance of PLMs trained on open-domain natural language.

Second, web queries are short and lack information. For example, in the CoNLL2003 \citep{tjong-kim-sang-de-meulder-2003-introduction} dataset that contains news articles, the average sequence length is 14.5. However, the average length of ads queries is only 3.9 on a self-collected dataset. This is problematic because the short web queries often lack semantic components for PLMs to make informed predictions. For example, in the natural language sentence ``Rabinovich is winding up his term as ambassador'', we can easily infer that ``\textit{Rabinovich}'' is a person's name. However, for a web query such as ``credit card square'', PLMs are unlikely to predict that "\textit{square}" is a brand.

% Second, web queries are short and lack information. In Table~\ref{tab:seq-len-dist}, we list the sequence length distribution of data from three datasets: two widely-used public datasets \citep{tjong-kim-sang-de-meulder-2003-introduction, pradhan-etal-2013-towards}, and a dataset comprising web search queries that we collected (termed AdsQuery). We see that sequences in AdsQuery are significantly shorter than natural language sequences such as news articles from CoNLL2003. This is problematic because the short web queries often lack semantic components for PLMs to make informed predictions. For example, in the natural language sentence ``Rabinovich is winding up his term as ambassador'', we can easily infer that ``\textit{Rabinovich}'' is a person's name. However, for a web query such as ``credit card square'', PLMs are unlikely to predict that "\textit{square}" is a brand.

The third problem is label scarcity. NER tasks demand token-level labels, requiring experienced and well-trained human annotators. As a result, domain-specific labeled data of good quality are limited \citep{jiang2021named, zhang2021queaco}. Existing works leverage weak supervision to tackle this issue. That is, instead of training human experts to accurately annotate data, automatic tools are used to generate noisy labels. For example, we can match the inputs to external knowledge bases \citep{liang2020bond, jiang2021named} or semantic rules \citep{yu2020fine, mukherjee2020uncertainty, awasthi2020learning}. However, labels generated by weak supervision can be extremely noisy, rendering the training process unstable \citep{zuo2021self}.

To tackle the above three challenges, we propose {\ours}, a knowledge-enhanced NER model for web-based ads queries. Specifically, we propose a knowledge enhancement framework that incorporates both \textit{model-free} and \textit{model-based} knowledge enhancement. We further propose a three-stage training framework to train {\ours} models.

We adopt three model-free knowledge enhancement methods:
(1) To address the domain shift issue, we collect large quantities of unlabeled web query data. The syntactic and semantic knowledge in these data can help models adapt to the advertisement domain.
(2) We also collect web search results to complement the lack of information in ads queries. Specifically, for each ads query, we retrieve several search results and keep the titles of these results. The web titles provide more context than the query, allowing models to better infer the role of each token (see Table~\ref{tab:web4ads-example} for examples).
(3) To alleviate the label scarcity problem, we collect weakly-labeled data from several sources. First, we resort to crowdsourcing platforms to collect inaccurate labels. Second, we leverage the Chain-of-Thoughts prompting \citep{wei2022chain} to automatically generate weak labels using large language models (LLMs) such as ChatGPT and GPT-4 \citep{openai2023gpt} (see Figure~\ref{fig:prompt} for an example). We remark that to the best of our knowledge, we are one of the first to augment data using LLMs for NER.
% Third, we adopt a translation-then-align approach to create labeled data for languages other than English. This is plausible because English data are much easier to collect than data in other languages.

In practice, we find that fine-tuning PLMs on NER tasks still faces severe overfitting. Therefore, we propose a model-based knowledge enhancement method based on adversarial regularization \citep{miyato2016adversarial, miyato2018virtual}. Specifically, during each training iteration, we find samples on which the model is likely to make erroneous predictions and augment the training data with such samples. The proposed augmentation technique improves model generalization by promoting prediction smoothness.

We train {\ours} models using a three-stage framework.
In \textbf{Stage I}, we continue pre-training \citep{gururangan2020don} a PLM on unlabeled web query data to inject domain knowledge.
In \textbf{Stage II}, we train the PLM from the previous stage on a large amount of weakly-labeled data.
In \textbf{Stage III}, we fine-tune the PLM from Stage II using model-based knowledge enhancement on a small amount of strongly-labeled data.
We note that similar multi-stage training methods have been shown to be effective in various tasks \citep{liang2020bond, yu2020fine, jiang2021named, zuo2021self}.

% \begin{table}[t]
% \centering
% \caption{Distribution of sequence length of natural language and ads queries. \textit{AdsQuery} is the dataset we collected.}
% \label{tab:seq-len-dist}
% \vspace{-0.1in}
% \begin{tabular}{l|ccccc|c}
% \toprule
% \multirow{2}{*}{\textbf{Dataset}} & \multicolumn{5}{c|}{\textbf{Percentile}} & \multirow{2}{*}{\textbf{Mean}} \\
% & \textbf{25\%} & \textbf{50\%} & \textbf{75\%} & \textbf{90\%} & \textbf{99\%} & \\ \midrule
% CoNLL2003 \citep{tjong-kim-sang-de-meulder-2003-introduction} & 6 & 10 & 22 & 32 & 46 & 14.5 \\
% OneNote5.0 \citep{pradhan-etal-2013-towards} & 8 & 14 & 23 & 34 & 59 & 17.3 \\ \hline
% AdsQuery & 3 & 4 & 5 & 6 & 10 & 3.9 \\
% \bottomrule
% \end{tabular}
% \end{table}

%% file: 0-data-augmentation.tex
\section{Model-Free Knowledge Enhancement}

%%%%%%%%%%%%%%%%%%%%%%%%%%%%%%%%%%%%%%%%%%%%%%%%%%%%%%%%%%%%
\subsection{Search Titles Complement Short Queries}

Web queries are inherently short, which presents a problem: the lack of semantic components in these queries often makes it difficult for pre-trained models to perform well. To address this issue, we propose augmenting web queries with search titles. Specifically, we retrieve search results from the search engine for each web query and keep the titles of the results. We refer to these titles as \textit{Web4Ads}.

\begin{table}[t]
\centering 
\caption{Two examples of web queries and the associated titles of search results (termed \textit{Web4Ads}).}
\label{tab:web4ads-example}
\begin{tabular}{l|l}
\toprule
\textbf{Query} & remote it support tools \\ \hline
\multirow{3}{*}{\textbf{Web4Ads}} & Best remote desktop software \\
& Best Tools to Easily Perform Remote Tech Support \\
& Best Remote Access Software \\ \toprule 
\textbf{Query} & credit card square \\ \hline
\multirow{3}{*}{\textbf{Web4Ads}} & Square: Solutions \& Tools to Grow Your Business \\
& Square Payments: Payment Processing \\
& Square Review: Fees, Complaints \\
\bottomrule
\end{tabular}
\end{table}

% \begin{figure}[t]
%     \centering
%     \includegraphics[width=0.8\linewidth]{figures/back-translation.png}
%     \caption{Illustration of translate-then-align. The black font indicates queries, and the red font indicates labels.}
%     \label{fig:back-translation}
% \end{figure}

Intuitively, the Web4Ads titles contain richer information than the original web queries, which can help the model better understand the role of each token in the query. Table~\ref{tab:web4ads-example} provides two examples. First, consider the query ``remote it support tools''. With only the web query, the model is likely to label the token ``it'' as \textit{other}. However, the Web4Ads titles associate ``it'' with entities such as ``tech'' and ``software''. Therefore, the model has a better chance of understanding that ``it'' is a \textit{product}. Second, for the query ``credit card square'', even human annotators have trouble recognizing that ``square'' is a company's name. However, after augmenting the Web4Ads titles, the model can infer that ``square'' is a \textit{brand} instead of a shape description.

During training, we concatenate a web query with its Web4Ads titles using a "[SEP]" token as the separator, e.g., an example is "Query [SEP] Web4Ads\textsubscript{1} [SEP] Web4Ads\textsubscript{2}".
We do not collect labels for the Web4Ads titles. Consequently, we do not compute the supervised loss for tokens that corresponds to the Web4Ads titles.
Empirically, our approach avoids introducing additional labeling burden and greatly improves model performance (see Section~\ref{sec:analysis} in the experiments for details).

\begin{figure}[t]
    \centering
    \includegraphics[width=0.6\linewidth]{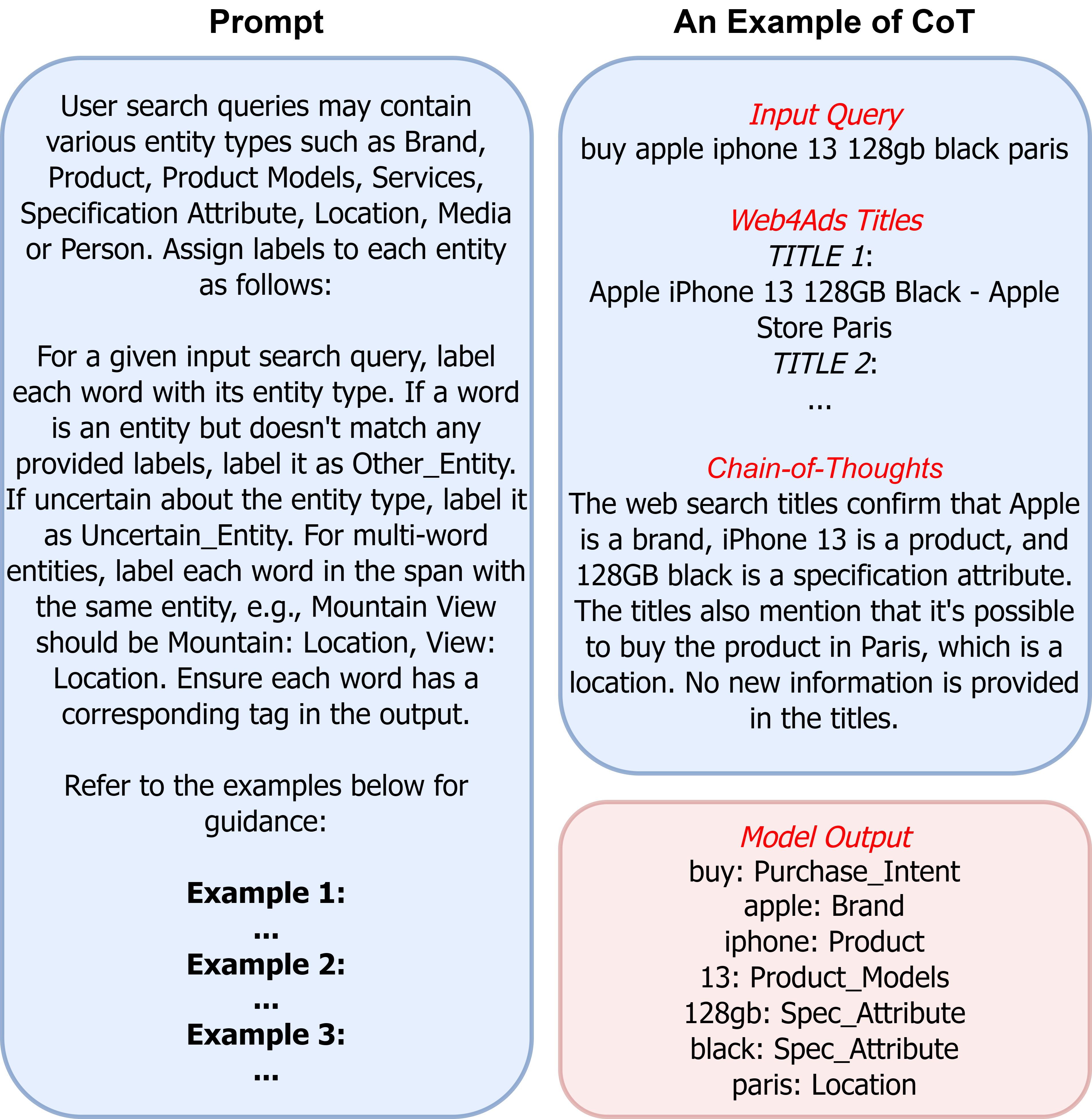}
    \caption{Prompt template for generating weak labels using large language models. The blue chunks are model inputs, and the red chunk contains model outputs.}
    \label{fig:prompt}
\end{figure}

%%%%%%%%%%%%%%%%%%%%%%%%%%%%%%%%%%%%%%%%%%%%%%%%%%%%%%%%%%%%
\subsection{Augmentation of In-Domain Data}

\noindent $\diamond$ \textbf{Unlabeled data}.
Web queries differ from natural language in that they often contain uncommon tokens (e.g., specification attributes and product models) and lack grammatical components such as verbs.
To address this domain shift issue, we collect large quantities of unlabeled web queries. 
The domain-specific semantic and syntactic knowledge in such data can be injected into the models.

\vspace{0.05in}
\noindent $\diamond$ \textbf{Weakly-labeled data}.
Experienced and well-trained human annotators are required to generate accurate token-level labels for NER tasks. As a result, strongly (or accurately) labeled data are often scarce in practice due to cost concerns. To address this issue, we leverage several sources to generate weakly-labeled data.

We collect crowdsourcing data. We remark that even though the data is human-annotated, the quality is often weak due to task difficulty and the absence of properly trained annotators.

We generate weakly-labeled data using large language models \citep{yoo2021gpt3mix, sahu2022data, liu2023first}. The development of models such as ChatGPT and GPT-4 \citep{openai2023gpt} has enabled weak-label generation with nearly no cost.
Specifically, we leverage Chain-of-Thoughts prompting \citep{wei2022chain}, where we use the Web4Ads titles as intermediate reasoning steps to guide the ``thinking'' of large language models (see Figure~\ref{fig:prompt} for an example).
We thoroughly investigate the effectiveness of several other prompting methods in Section~\ref{sec:prompt}.

% For multilingual data, we propose a translation-then-align method to generate weak labels. In general, labeled data are easier to collect for queries in English than in other languages such as French and German. For a labeled query in English, we first translate the query into another language using off-the-shelf translation models. Then, we align the translated tokens with the original ones to obtain labels for the translated query. Figure~\ref{fig:back-translation} illustrates the translation-then-align procedure.

%% file: 0-adv-augmentation.tex
\section{Model-Based Knowledge Enhancement}

\begin{figure}[t]
    \centering
    \includegraphics[width=0.55\linewidth]{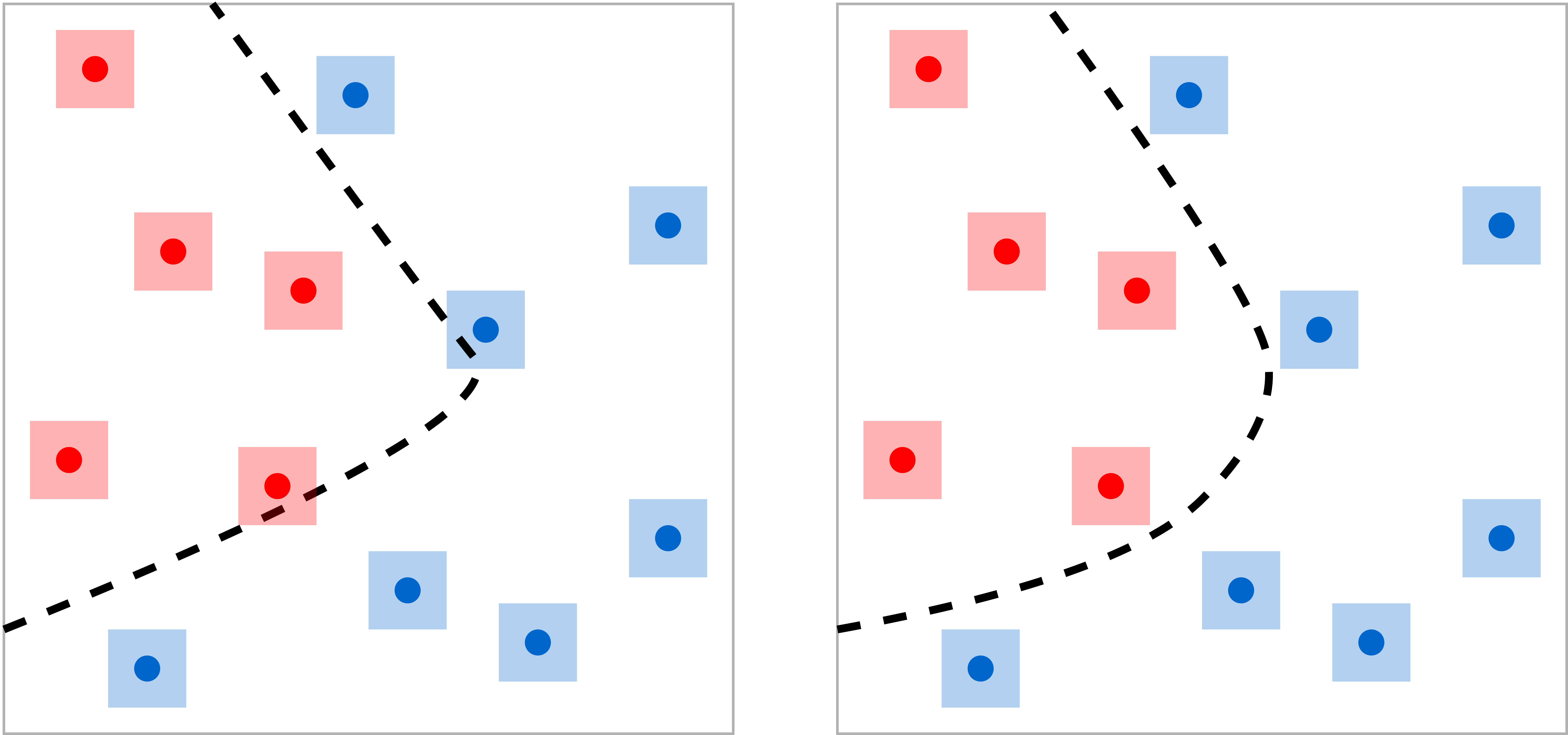}
    \caption{Illustration of decision boundaries without (left) and with (right) adversarial knowledge enhancement. A solid circle indicates a labeled sample, and the square around it indicates its neighborhood. The red and blue colors indicate two difference classes of samples, and the dashed lines are decision boundaries.}
    \label{fig:adv-demo}
\end{figure}

Fine-tuning pre-trained language models require large quantities of labeled data, which are often unavailable. Even though the proposed model-free knowledge enhancement methods can partially alleviate this issue, in practice, extensive hyper-parameter tuning is still needed to avoid overfitting.
We propose a model-based data augmentation method based on adversarial regularization \citep{miyato2016adversarial, miyato2018virtual} to reduce over-fitting.

%%%%%%%%%%%%%%%%%%%%%%%%%%%%%%%%%%%%%%%%%%%%%%%%%%%%%%%%%%%%
\subsection{Virtual Data Augmentation}
\label{sec:virtual-aug}

Given a single datum, if we perturb it by a small noise, the model's prediction should not change. Such a smoothness assumption promotes the model's generalization performance \citep{miyato2018virtual}.
Therefore, for each labeled sample, we can augment the training dataset by generating virtual data in its neighborhood \citep{awasthi2020learning}.

Figure~\ref{fig:adv-demo} illustrates the idea. From Figure~\ref{fig:adv-demo} (left), we see that without augmentation, some data are very close to the decision boundary. Therefore, a small perturbation to the input data may result in a substantial change to the model's prediction.
Virtual data augmentation encourages the model to make consistent predictions under small perturbations. For example, in Figure~\ref{fig:adv-demo}, if the model predicts a sample as ``red'', then it should classify all virtually generated data around the sample (i.e., data in the sample's neighborhood) as ``red''.
In this way, the decision boundary becomes smoother, which improves model generalization.

%%%%%%%%%%%%%%%%%%%%%%%%%%%%%%%%%%%%%%%%%%%%%%%%%%%%%%%%%%%%
\subsection{Model-Based Adversarial Data Augmentation}

\begin{figure}[t]
    \centering
    \begin{subfigure}{0.25\linewidth}
         \centering
         \includegraphics[width=\linewidth]{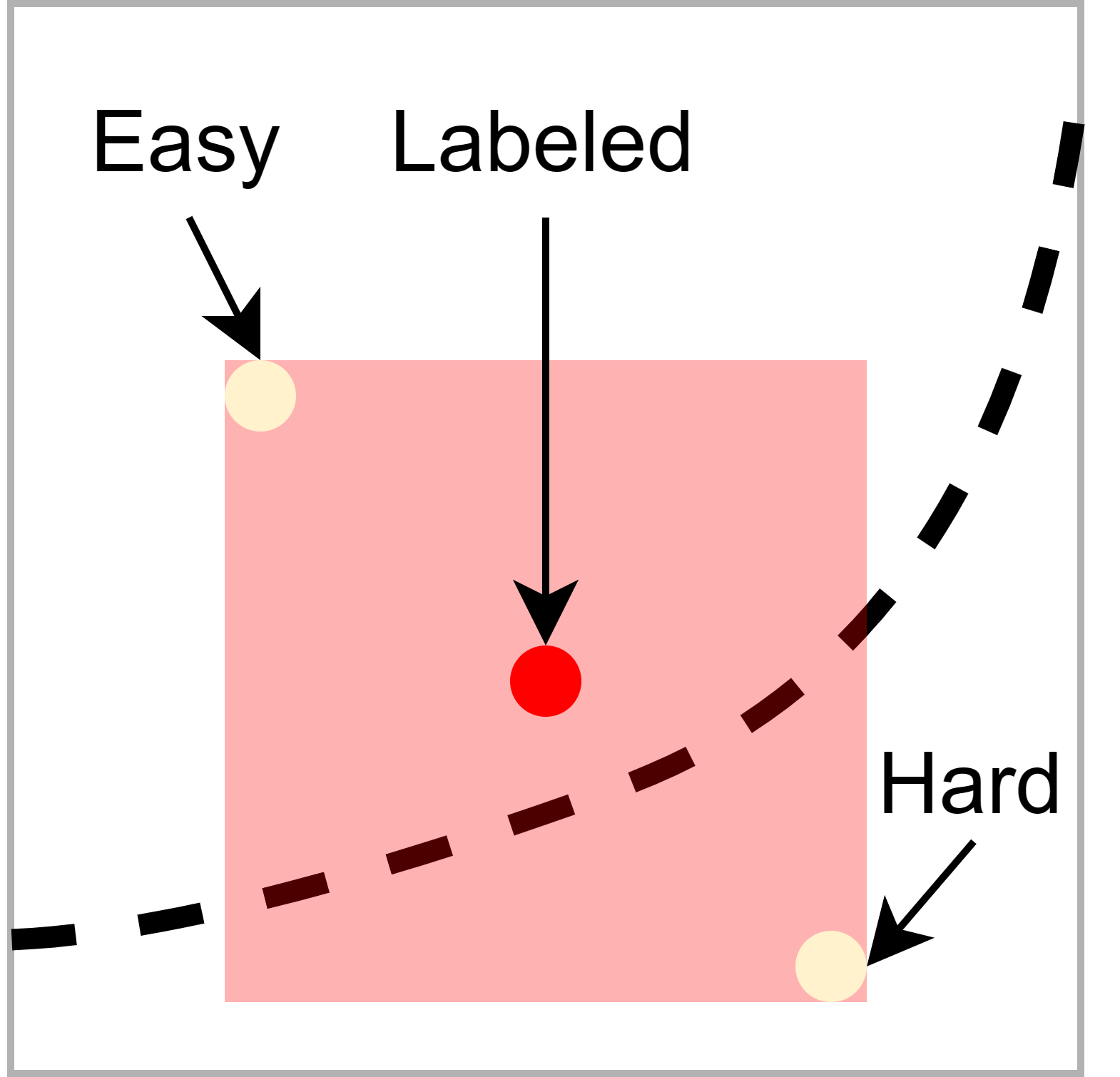}
         \caption{Illustration of hard and easy virtual samples.}
         \label{fig:hard-and-easy}
     \end{subfigure} \hspace{0.3in}
     \begin{subfigure}{0.25\linewidth}
         \centering
         \includegraphics[width=\linewidth]{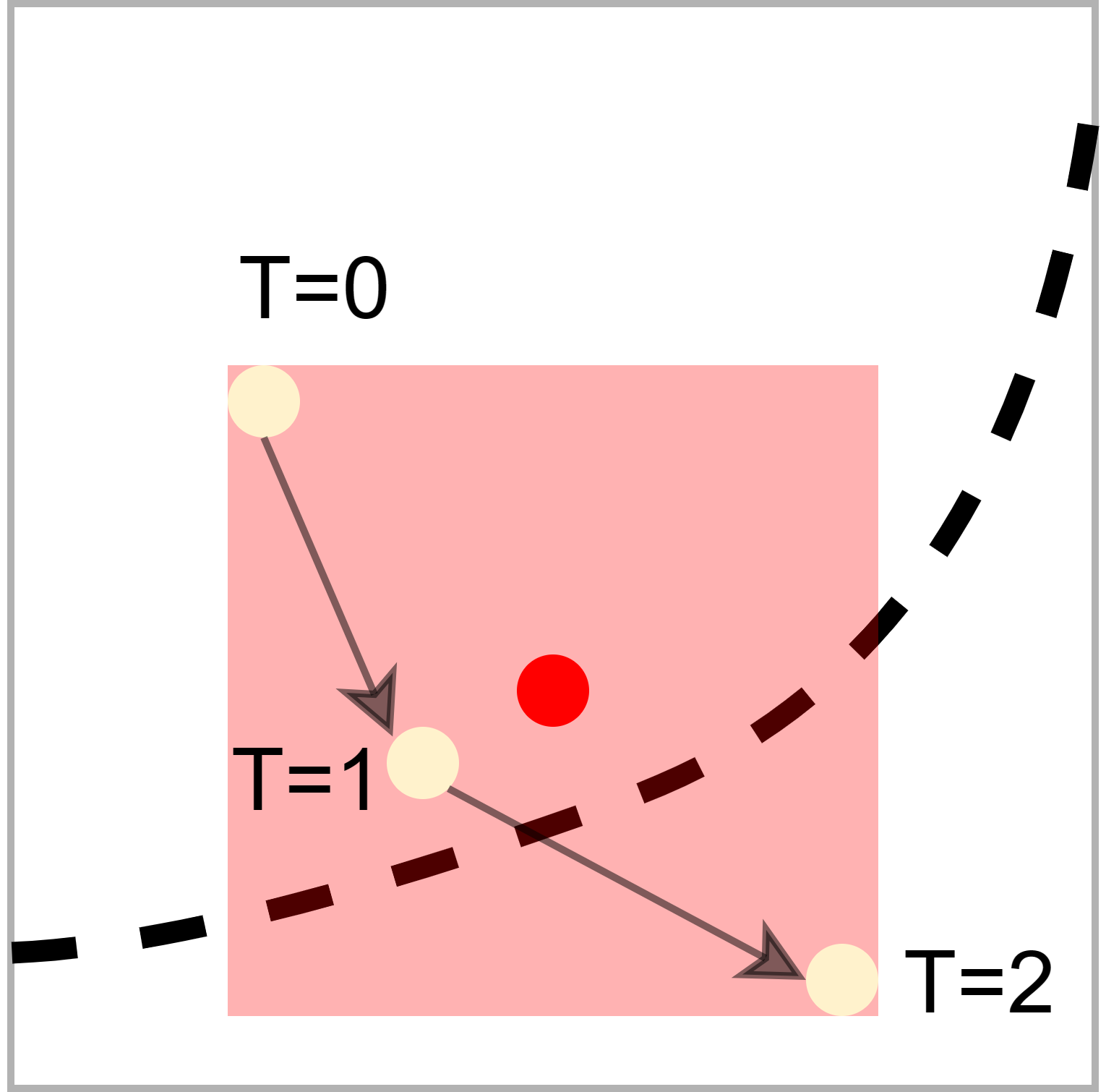}
         \caption{Illustration of finding hard virtual samples.}
         \label{fig:sample-update}
     \end{subfigure}
    \caption{Illustration of virtual samples. The red solid circle indicates a labeled sample, and the square indicates its neighborhood. The light yellow circles are virtual samples.}
\end{figure}

There are infinitely many virtual data in the neighborhood of a labeled sample. However, not all of the virtual samples are of the same ``difficulty''.
For example, in Figure~\ref{fig:hard-and-easy}, the model correctly classifies the labeled sample (the red circle). In this case, augmenting the data with the \textit{easy} virtual sample (the yellow circle on the top left) will not bring any improvement since the model can already correctly classify it. On the other hand, the model predicts the \textit{hard} virtual sample (the yellow circle on the bottom right) to a different class, which violates our assumption that neighboring data should have the same label. Therefore, augmenting this hard virtual sample will benefit model generalization (see Figure~\ref{fig:adv-demo}).

We find the hard virtual samples via adversarial training.
Concretely, denote $f(x,\theta)$ a neural network parameterized by $\theta$, where $x$ is the input. We note that $x$ is continuous and resides in the embedding space. For example, for NER, $x$ is the continuous representation after the embedding layers. 
Then, We find the hard virtual samples by optimizing the following objective
\begin{align} \label{eq:loss-adv}
    &\max_{\norm{\delta} \leq \epsilon} \ell_v(x, \delta, \theta), \\
    &\text{where}\ \ell_v (x, \delta, \theta) = \mathrm{SymKL} \left( f(x,\theta), f(x+\delta, \theta) \right). \notag
\end{align}
Here, $\mathrm{SymKL}(P, Q) = \frac{1}{2} \left(\mathrm{KL}(P||Q) + \mathrm{KL}(Q||P) \right)$ is the symmetric KL-divergence between two probability distributions; $\norm{\cdot}$ is taken as the $\ell_2$ norm or the $\ell_{\infty}$ norm; and $\epsilon$ is a pre-defined radius of the perturbations.
The loss $\ell_v$ measures discrepancy between the model's predictions given the clean data $x$ and the perturbed data $x+\delta$. Thus, by maximizing $\ell_v$, we can find virtual samples on which the model easily make erroneous predictions (e.g., the hard virtual sample in Figure~\ref{fig:hard-and-easy}).

\begin{figure*}[t]
    \centering
    \includegraphics[width=1.0\textwidth]{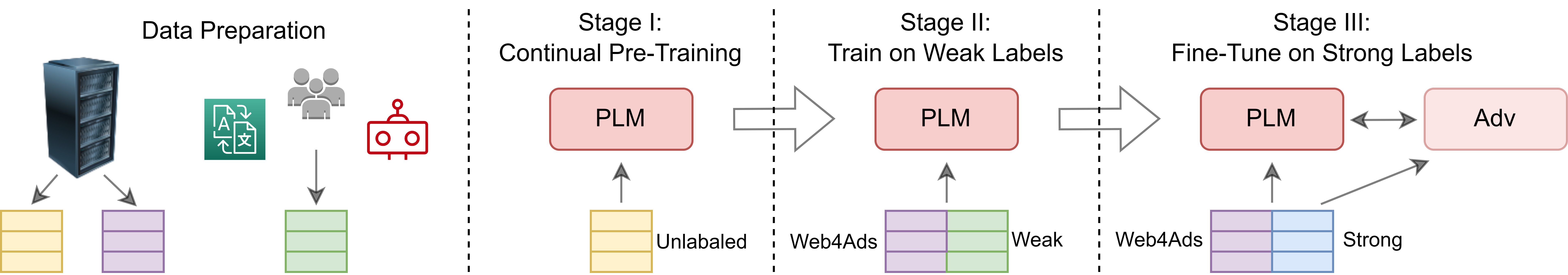}
    \caption{Overall framework of {\ours}.  In data preparation, unlabeled and Web4Ads data are retrieved from search engines; and weak labels are generated from multiple sources. Strongly-labeled data in Stage III are annotated by human experts.}
    \label{fig:arch}
\end{figure*}

In practice, we solve Equation~\ref{eq:loss-adv} using projected gradient ascent \citep{madry2017towards}. That is, we adopt the following update rule:
\begin{align*}
    \delta_{k+1} = \Pi \left( \delta_k + \eta \frac{\nabla_\delta \ell_v(x, \delta_k, \theta)}{\norm{\nabla_\delta \ell_v}_2} \right), \
    \text{where}\ \delta_0 \sim \Pi \left( \mathcal{N}(\mathbf{0}, \mathbf{I}) \right).
\end{align*}
Here, $\Pi(\cdot)$ is the projection operator onto the $\norm{\cdot}$ ball, and $\eta$ is the learning rate.
Figure~\ref{fig:sample-update} illustrates finding hard virtual samples. At first ($T=0$), we randomly initialize a perturbation. Then, we update the perturbation by moving it towards the decision boundary ($T=1$). Finally ($T=2$) the model makes a wrong prediction, and we deem the resulting virtual sample \textit{hard}.

To leverage the hard virtual samples during training, we optimize
\begin{align*}
    \min_\theta \frac{1}{n}\sum_{i=1}^n \ell(f(x_i, \theta), y_i)
    + \max_{\norm{\delta_i} \leq \epsilon} \ell_v(x_i, \delta_i, \theta),
\end{align*}
where $y_i$ is the ground-truth label corresponding to $x_i$, and $\ell$ is the cross-entropy loss.

We note that virtual data augmentation in Section~\ref{sec:virtual-aug} has been shown to improve model generalization \citep{aghajanyan2020better}. However, in practice, adversarial data augmentation works better in terms of both model performance and training stability \citep{zuo2021arch}.

%% file: 0-training.tex
\section{Training of {\ours}}

We propose a three-stage training framework to train {\ours} models. The framework is illustrated in Figure~\ref{fig:arch}.

In \textbf{Stage I}, we continue pre-training a pre-trained language model (PLM) on the collected unlabeled web queries. Specifically, we train the PLM using self-supervision objectives such as masked language modeling \citep{devlin2018bert}. This can effectively address the domain shift issue by adapting the PLM to the advertisement domain. 

In \textbf{Stage II}, we train the PLM from the previous stage on a large amount of weakly-labeled data, augmented with Web4Ads titles. To prevent the PLM from overfitting to the noise in the weak labels, we adopt an early-stopping strategy \citep{dodge2020fine}. We note that other weakly-supervised learning methods such as contrastive learning \citep{yu2020fine} and confidence regularization \citep{pereyra2017regularizing} can be applied in this stage.

In \textbf{Stage III}, we fine-tune the PLM from Stage II on a small amount of strongly-labeled data, augmented with Web4Ads titles. We also apply model-based adversarial data augmentation in this stage. We note that the strongly-labeled data are annotated by human experts, and therefore the amount of them is magnitudes smaller than the weakly-labeled data.

%% file: 0-data-preparation.tex
\section{Data Preparation}
\label{sec:prompt}

%%%%%%%%%%%%%%%%%%%%%%%%%%%%%%%%%%%%%%%%%%%%%%%%%%%%%%%%%%%%
\subsection{Data Overview}

Table~\ref{tab:dataset} summarizes the collected data. Specifically, we use unlabeled web search queries collected from search engines for continual pre-training. 
We also collect strongly-labeled data, where the labels are annotated by human experts. However, the quantity of such data is limited because of cost concerns.

Additionally, we collect crowdsourcing data for queries in English. And for other languages (e.g., German and French), we generate weak labels using LLMs (see Section~\ref{sec:wek-llm} for details). We do not collect crowdsourcing data in German and French because of the lack of human annotators.

\begin{table}[t]
\centering
\caption{Dataset statistics. In all the experiments, we report results on the test set sampled from strongly-labeled data.}
\label{tab:dataset}
\begin{tabular}{l|c|cc|c}
\toprule
\textbf{Language} & \textbf{Type} & \textbf{\#Train} & \textbf{\#Test} & \textbf{\#Classes} \\ \midrule
\multirow{3}{*}{En} & unlabeled & 998.7M & -- & \multirow{3}{*}{11} \\ 
& crowdsource & 2.6M & -- & \\
& strongly-labeled & 167.4k & 2.0k & \\ \hline
\multirow{3}{*}{De} & unlabeled & 120.8M & -- & \multirow{3}{*}{11} \\ 
& LLM & 100.0k & -- & \\
& strongly-labeled & 51.5k & 2.2k & \\ \hline
\multirow{3}{*}{Fr} & unlabeled & 91.5M & -- & \multirow{3}{*}{11} \\ 
& LLM & 100.0k & -- & \\
& strongly-labeled & 50.0k & 2.2k & \\ \hline
\end{tabular}
\end{table}

\begin{table}[t]
\centering
\caption{Comparison of different prompting methods. Here, ``+'' denotes positive examples, ``-'' denotes negative ones, and ``(+,-)'' denotes positive examples followed by negative ones.}
\label{tab:prompts}
\begin{tabular}{l|c|c|c|c}
\toprule
\textbf{Method} & \textbf{Few Shots} & \textbf{Embedding} & \textbf{Web4Ads} & \textbf{F1} \\ 
\midrule
Prompting & No & No & No & 0.62 \\
Demo. & Yes & No & No & 0.72 \\
Dyna. Demo. & Yes & EASE (+) & No & 0.67 \\
Dyna. Demo. & Yes & EASE (+,-) & No & 0.69 \\
Dyna. Demo. & Yes & EASE (-,+) & No & 0.70 \\
Dyna. Demo. & Yes & SBERT (+) & No & 0.68 \\
CoT & Yes & No & top 3 & \textbf{0.80} \\
\bottomrule
\end{tabular}
\end{table}

%%%%%%%%%%%%%%%%%%%%%%%%%%%%%%%%%%%%%%%%%%%%%%%%%%%%%%%%%%%%
\subsection{Weak Labels from Large Language Models}
\label{sec:wek-llm}

Large language models such as ChatGPT excels at straightforward tasks such as text classification  \citep{ding2022gpt, chiang2023can, wang2023chatgpt, wu2023large}. However, they encounter limitations when dealing with more intricate problems such as NER.

We explore a variety of methods aimed at enhancing prompts for label generation.
\begin{itemize}
\item[$\diamond$] \textbf{Prompting}. In this approach, we only use the prompt (without examples) in Figure~\ref{fig:prompt} (left) to generate labels.
\item[$\diamond$] \textbf{Demonstration}. We utilize a few-shot demonstrations approach \citep{liu2023pre, xie2021explanation} with fixed instances (Figure~\ref{fig:prompt} left). Specifically, we select three fixed examples that ensure full coverage across all categories.
\item[$\diamond$] \textbf{Dynamic demonstration}. In this case, the demonstrations are dynamically retrieved. Specifically, we retrieve three positive and/or three negative examples using two existing embedding models: SBERT \citep{reimers2019sentence} and EASE \citep{nishikawa2022ease}. This approach facilitates contrastive in-context learning.
\item[$\diamond$] \textbf{Chain-of-Thoughts}. We enhance the prompt using web information via Chain-of-Thoughts prompting \citep{wei2022chain}. Specifically, we augment the Web4Ads titles to intermediate reasoning steps (Figure~\ref{fig:prompt} right).
\end{itemize}
To evaluate the prompting methods, we generate labels on strongly-labeled test sets (Table~\ref{tab:dataset}). Then, we calculate the F1 score of the ``\textit{Brand}'' category (there are 11 categories in total) since it plays a crucial role in downstream ads-related tasks.

As shown in Table~\ref{tab:prompts}, vanilla \textit{Prompting} yields a brand F1 score of 0.62. By employing few-shot demonstrations (\textit{Demo}), the F1 score increases to 0.72.
However, we observe a performance degradation when employing \textit{Dynamic Demonstration} methods with EASE and SBERT. This is because both embedding models are unsuitable for NER. In particular, SBERT does not incorporate entity-related information, and EASE falls short in capturing the relationships between entities and their contexts. 
To better leverage the entity information and the augmented web titles, we integrate a Chain-of-Thoughts layer, which effectively guides the LLMs to extract valuable information from Web4Ads titles. This approach results in a substantial increase in the F1 score, raising it to 0.80.

We remark that in reality, usage of LLMs in online deployment is limited due to latency constraints. For example, our {\ours} model can output token labels within milliseconds, but LLMs require several hundred milliseconds or even seconds.

%% file: 0-experiments.tex
\section{Experiments}

\begin{table}[t]
\centering
\caption{Experimental results on English queries. We report the overall F1 score and the F1 score that correspond to the ``Brand'' category. The best results are shown in \textbf{bold}.}
\label{tab:main-results-en}
\begin{tabular}{l|cc|cc}
\toprule \small
\textbf{English} & \textbf{Weak Data} & \textbf{Weak Labels} & \textbf{Brand} & \textbf{Overall} \\ \midrule
\multicolumn{3}{l}{\textbf{Direct training}} \\
BERT & \xmark & \xmark & 80.77 & 74.29 \\ \hline
\multicolumn{3}{l}{\textbf{Semi-supervised baselines}} \\
Self-Training & \cmark & \xmark & 81.20 & 74.76 \\
DRIFT & \cmark & \xmark & 81.61 & 75.12 \\
VAT & \cmark & \xmark & 82.04 & 75.86 \\ \hline
\multicolumn{3}{l}{\textbf{Weakly-supervised baselines}} \\
COSINE & \cmark & \cmark & 81.34 & 74.69 \\
NEEDLE & \cmark & \cmark & 82.27 & 76.33 \\ \hline
\multicolumn{3}{l}{\textbf{Ours}} \\
{\ours} & \cmark & \cmark & \textbf{83.45} & \textbf{77.94} \\
\bottomrule
\end{tabular}
\end{table}

%%%%%%%%%%%%%%%%%%%%%%%%%%%%%%%%%%%%%%%%%%%%%%%%%%%%%%%%%%%%
\subsection{Baselines}

We implement all the models using \textit{PyTorch} \citep{paszke2019pytorch} and the \textit{Huggingface Transformers} \citep{wolf2020transformers} code-base.
In the experiments, we fine-tune a BERT-base model for English data, and we fine-tune a multilingual version of BERT-base for data in other languages. To facilitate fair comparisons, we conduct unsupervised continual pre-training (i.e., \textit{Stage I} in our framework) for all models.

We compare {\ours} with several weakly-supervised and semi-supervised learning baselines.
We note that for semi-supervised learning baselines, we first fine-tune a BERT model on the strongly-labeled data, and we treat the resulting model as the ``teacher''. We do not use the weak labels in the weakly-labeled data, and instead we use the teacher model to generate pseudo-labels for these data.

\noindent
$\diamond$ \textit{BERT} \citep{devlin2018bert} is where we only train on the strongly-labeled data, i.e., without weakly-labeled data.

\noindent
$\diamond$ \textit{Self-Training} \citep{rosenberg2005semi, lee2013pseudo} is a classic semi-supervised learning approach. In Self-Training, we simultaneously maintain a teacher and a student model. The teacher generates pseudo-labels, on which the student is trained. The two models are updated alternatingly.

\noindent
$\diamond$ \textit{DRIFT} \citep{zuo2021self} adopt a differentiable self-training approach. The method stabilizes conventional self-training by formulating the mean-teacher framework as a Stackelberg game. 

\noindent
$\diamond$ \textit{VAT} \citep{miyato2016adversarial, miyato2018virtual} is a semi-supervised learning method that employs adversarial training. Specifically, VAT regularizes model training by penalizing the divergence between model predictions on clean and perturbed unlabeled data.

\noindent
$\diamond$ \textit{COSINE} \citep{yu2020fine} is a weakly-supervised learning framework that can efficiently leverage noisily-labeled data. The framework adopts token-level contrastive learning, and also integrates the power of confidence-based sample re-weighting and regularization.

\noindent
$\diamond$ \textit{NEEDLE} \citep{jiang2021named} is a NER framework that use small strongly-labeled data and large weakly-labeled data. The framework proposes a noise-aware loss function for weakly-supervised learning.

%%%%%%%%%%%%%%%%%%%%%%%%%%%%%%%%%%%%%%%%%%%%%%%%%%%%%%%%%%%%
\subsection{Main Results}

% \newcolumntype{C}{@{\hskip3pt}c@{\hskip3pt}}
\begin{table}[t]
\centering
\caption{Results on German and French queries. We report the overall F1 score and the F1 score that correspond to the ``Brand'' category. The best results are shown in \textbf{bold}.}
\label{tab:main-results-defr}
\begin{tabular}{l|cccccc}
\toprule
& \multicolumn{2}{c}{\textbf{German}} & \multicolumn{2}{c}{\textbf{French}} & \multicolumn{2}{c}{\textbf{Average}} \\
& \textbf{Brand} & \textbf{Overall} & \textbf{Brand} & \textbf{Overall} & \textbf{Brand} & \textbf{Overall} \\ \midrule
BERT & 67.11 & 58.21 & 69.58 & 61.45 & 68.35 & 59.83 \\ 
VAT & 67.41 & 58.32 & 69.62 & 61.45 & 68.52 & 59.89 \\
NEEDLE & 67.80 & 58.41 & 69.91 & 61.57 & 68.86 & 59.99 \\ \hline
{\ours} & \textbf{68.59} & \textbf{59.24} & \textbf{70.57} & \textbf{62.63} & \textbf{69.58} & \textbf{60.94} \\
\bottomrule
\end{tabular}
\end{table}

Table~\ref{tab:main-results-en} summarizes results on English queries; and Table~\ref{tab:main-results-defr} summarizes results on German and French queries. From the results, we see that {\ours} performs significantly better than all the semi-supervised and weakly-supervised baselines.

We note that in semi-supervised learning, we do not use the weak labels from crowdsourcing platforms and LLMs, which is different from weakly-supervised learning. Because of this, we see that the best-performing semi-supervised learning baseline (VAT) behaves worse than the best-performing weakly-supervised learning baseline (NEEDLE) in terms of both Brand F1 score and overall F1 score.

Also, we note that in weakly-supervised learning, COSINE adopts a different training framework, where in the last stage the model is trained on both weakly-labeled and strongly-labeled data. In contrast, in {\ours} and NEEDLE, the model is first trained on weakly-labeled data and then fine-tuned on strongly-labeled data. The results in Table~\ref{tab:main-results-en} and Table~\ref{tab:main-results-defr} indicate that the weak-then-strong training approach is better than training on both weakly-labeled and strongly-labeled data.

Even though {\ours} and NEEDLE employ a similar training framework, we see that performance of {\ours} is significantly better. In NEEDLE, vanilla training (i.e., without data augmentation or modification to the loss function) is used in final fine-tuning on strongly-labeled data. However, because strongly-labeled data are limited, the model can easily overfit to the noise. Our method can reduce outfitting and improve model generalization via adversarial data augmentation.

%%%%%%%%%%%%%%%%%%%%%%%%%%%%%%%%%%%%%%%%%%%%%%%%%%%%%%%%%%%%
\subsection{Online Deployment}
The described system has been deployed to Microsoft Bing Ads for approximately one year. During this period, {\ours} processes more than one billion queries daily. 
Compared with the previous NER system, {\ours} increases revenue by 0.7\%, increases Brand detection F1 score by 4.7\%, and increases overall entity detection F1 score by 3.2\%.

%%%%%%%%%%%%%%%%%%%%%%%%%%%%%%%%%%%%%%%%%%%%%%%%%%%%%%%%%%%%
\subsection{Analysis}
\label{sec:analysis}

\noindent $\diamond$
\textbf{Effectiveness of continual pre-training}.
Recall that in Stage I of {\ours}, we continue pre-train a BERT model on unlabeled web query data to inject domain knowledge. Table~\ref{tab:pre-train} summarizes the results. We see that continual pre-training improves model performance for both vanilla training and {\ours}. For example, without pretraining, {\ours} yields a 61.05 overall F1 score on French queries. And after integrating pre-training, the overall F1 score increases to 62.63 (+1.58).

\begin{table}[h]
\vspace{-0.1in}
\centering
\caption{Experimental results with and without continual pre-training. For ``BERT'', we directly fine-tune the model on strongly-labeled data without using weak labels.}
\label{tab:pre-train}
\vspace{-0.1in}
\begin{tabular}{ll|cc|cc}
\toprule
& & \multicolumn{2}{c|}{\textbf{BERT}} & \multicolumn{2}{c}{\textbf{{\ours}}} \\
& & \textbf{w/o} & \textbf{w/} & \textbf{w/o} & \textbf{w/} \\ \midrule
\multirow{2}{*}{English} & Brand & 79.71 & 80.77 & 82.29 & \textbf{83.45} \\
& Overall & 73.23 & 74.29 & 77.41 & \textbf{77.94} \\ \hline
\multirow{2}{*}{German} & Brand & 65.87 & 67.11 & 67.52 & \textbf{68.59} \\
& Overall & 57.08 & 58.21 & 57.73 & \textbf{59.24} \\ \hline
\multirow{2}{*}{French} & Brand & 68.56 & 69.58 & 69.16 & \textbf{70.57} \\
& Overall & 60.51 & 61.45 & 61.05 & \textbf{62.63} \\
\bottomrule
\end{tabular}
\end{table}

% \vspace{0.1in}
\noindent $\diamond$
\textbf{Effectiveness of adversarial data augmentation}.
Table~\ref{tab:adv} demonstrates model performance with and without adversarial data augmentation. First, we see that even without augmentation, performance of {\ours} is significantly better than vanilla training.
Second, we see that adversarial data augmentation and virtual data augmentation can indeed improve model performance by reducing overfitting. 
Third, notice that adversarial data augmentation performs better than virtual data augmentation (Section~\ref{sec:virtual-aug}) because of its ability to find ``hard'' virtual samples.

\begin{table}[h]
\centering
\caption{Experimental results with and without adversarial data augmentation. For ``BERT'', we directly fine-tune the model on strongly-labeled data without using weak labels. For ``virtual'', we use virtual data augmentation in Section~\ref{sec:virtual-aug}.}
\label{tab:adv}
\begin{tabular}{l|cccccc}
\toprule
& \multicolumn{2}{c}{\textbf{English}} & \multicolumn{2}{c}{\textbf{German}} & \multicolumn{2}{c}{\textbf{French}} \\
& \textbf{Brand} & \textbf{Overall} & \textbf{Brand} & \textbf{Overall} & \textbf{Brand} & \textbf{Overall} \\ \midrule
BERT & 80.77 & 74.29 & 67.11 & 58.21 & 69.58 & 61.45 \\ \hline
w/o aug. & \textbf{84.08} & 77.10 & 67.69 & 58.42 & 70.47 & 61.59 \\
virtual aug. & 83.47 & 77.62 & 67.72 & 58.72 & 70.32 & 62.13 \\
{\ours} & 83.45 & \textbf{77.94} & \textbf{68.59} & \textbf{59.24} & \textbf{70.57} & \textbf{62.63} \\ 
\bottomrule
\end{tabular}
\end{table}

\begin{figure}[h]
\centering
\includegraphics[width=0.4\linewidth]{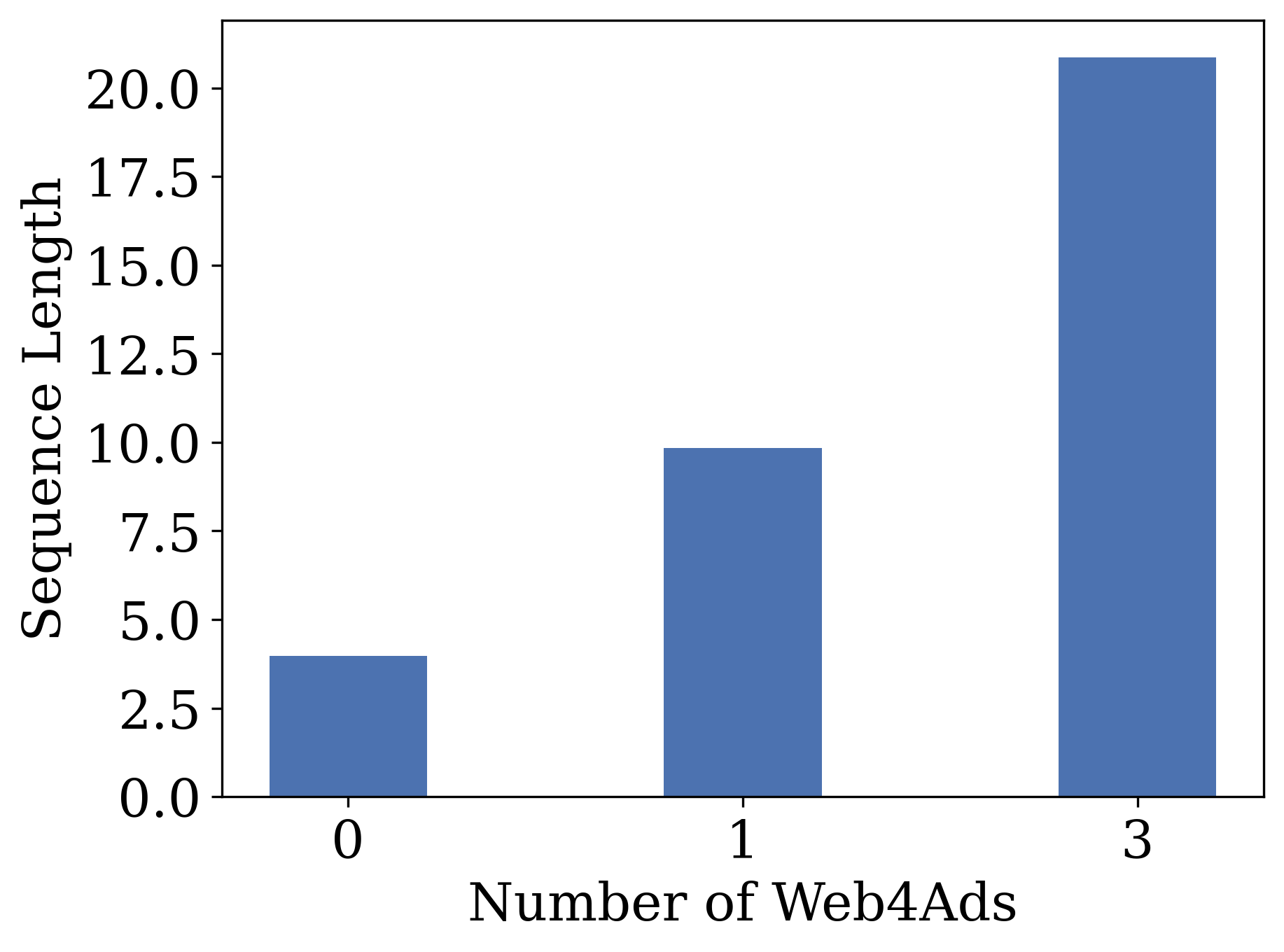} \hspace{0.3in}
\includegraphics[width=0.38\linewidth]{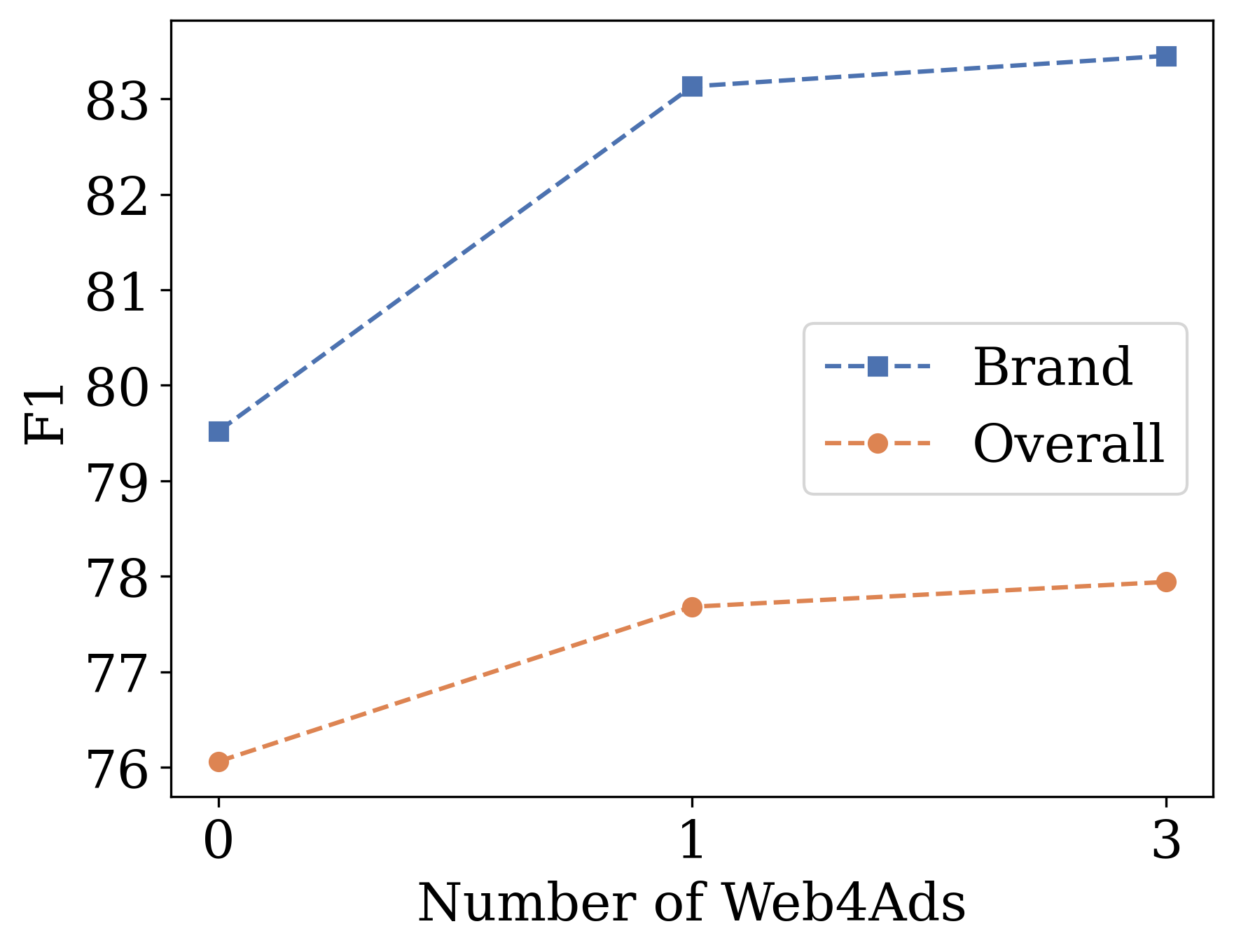}
\caption{Effectiveness of Web4Ads. Left: average sequence length; Right: model performance on English queries.}
\label{fig:web4ads}
\end{figure}

% \vspace{0.1in}
\noindent $\diamond$
\textbf{Effectiveness of Web4Ads}.
In {\ours}, we complement the short queries with Web4Ads titles. We verify the effectiveness of such an approach in Figure~\ref{fig:web4ads}. In Figure~\ref{fig:web4ads} (left), we plot the average sequence length with and without Web4Ads titles. We see that the average length of web queries is only 3.9, such that the queries alone might not be informative. After integrating the Web4Ads titles, the average length increases to 20.8. Therefore, the augmented queries contain richer information. As shown in Figure~\ref{fig:web4ads} (right), compared with only using the web queries, model performance significantly increases when we add one Web4Ads title. Also, model performance continues to increase when we further increase the number of Web4Ads titles.

%% file: 0-related-works.tex
\section{Related Works}

$\diamond$ \textbf{Weakly-supervised learning}.
In weakly-supervised learning, the labels are noisy and incomplete.
For example, we can obtain weak labels by writing semantic rules \citep{yu2020fine, mukherjee2020uncertainty, awasthi2020learning} or match the data to external knowledge bases \citep{liang2020bond, jiang2021named}. 
Existing methods aim to denoise the labels by using, for example, soft pseudo-labels \citep{xie2016unsupervised, xie2020unsupervised, meng2020text, liang2020bond, yu2020fine, zuo2021self}, confidence regularization \citep{pereyra2017regularizing, jiang2021named}, and labeling-function aggregation \citep{ratner2017snorkel, varma2018snuba, lison2020named}.

% \vspace{0.05in}
\noindent
$\diamond$ \textbf{Adversarial regularization for natural language processing}.
Adversarial training was originally proposed in computer vision \citep{szegedy2013intriguing, goodfellow2014explaining, madry2017towards}, with the goal of training classifiers that are robust to adversarial input images.
However, in natural language processing, the goal of adversarial training to to leverage its regularization effect to improve model generalization \citep{raghunathan2020understanding}.
Many works surrounding the efficiency \citep{zhu2019freelb, shafahi2019adversarial, aghajanyan2020better, wu2021r, zuo2021arch} and effectiveness \citep{cheng2020posterior, liu2020adversarial, zuo2021adversarial} of adversarial regularization have been proposed.

%% file: 0-conclusion.tex
\section{Conclusion and Discussion}

We propose {\ours}, a knowledge-enhanced NER model for web-based ads queries. {\ours} leverages both model-free and model-based knowledge enhancement methods. For model-free approaches, we collect unlabeled web queries to inject domain knowledge, and we collect web search titles to complement the short queries. Additionally, we generate weak labels using large language models such as ChatGPT. For model-based approaches, we employ a model-dependent augmentation methods based on adversarial training. Extensive experiments in various NER tasks demonstrate the effectiveness of {\ours}.

In {\ours}, we adopt a three-stage training framework. In the second stage, we train the model on weakly-labeled data without any modification to the loss function. This is different from previous approaches, e.g., \citep{yu2020fine} uses contrastive learning to suppress noise in weak labels. In practice, we observe only marginal differences when incorporating techniques such as contrastive learning and confidence regularization. We attribute this to the good coverage and quality of weak labels generated by LLMs. We leave further explorations on using LLMs to generate labels as future works.